# Intelligent Techniques for Resolving Conflicts of Knowledge in Multi-Agent Decision Support Systems

Khaled M. Khalil, M. Abdel-Aziz, Taymour T. Nazmy, Abdel-Badeeh M. Salem

*Faculty of Computer and Information Science Ain shams University Cairo, Egypt*

kmkmohamed@gmail.com
mhaziz67@gmail.com
ntaymoor@yahoo.com
abmsalem@yahoo.com

*Abstract*— The biggest challenge for agents' collaboration in Decision Support Systems is resolving possible conflicts of knowledge. When coordinating activities, either in a cooperative or a competitive environment, conflicts may arise and three basic strategies to solve these conflicts are by means of negotiation, mediation and arbitration. Following these strategies; different intelligent techniques developed for knowledge conflict resolution. This paper focuses on some of the key intelligent techniques for conflict resolution in Multi-Agent Decision Support Systems. It focuses on the part of agents' knowledge conflicts and it discusses seven techniques: Bayesian Network, Case Based Reasoning, Expert Systems, Fuzzy Systems, Genetic Algorithms, Ontological, and Searching based techniques. These techniques show how different technologies in the area of intelligent agents can be combined to solve this real-world decision support problem.

## I. INTRODUCTION

Decision Support Systems are software applications that have been used over the last few decades to provide support for many structured and unstructured problems such as Strategic Planning, Investment Planning, Human Resources Management, and Help Desk Automation. Decision Support Systems components such as Knowledge Management systems, Model Management systems and Data Management systems aid humans in making better decisions by incorporating previous knowledge and information about the domain. For example, in Strategic Planning, Decisions Support Systems plan and utilize decisions of production capacity [1, 2]. While in Investment Planning, Decision Support Systems provide justification and planning of the organization investments [3, 4]. In Human Resources Management, Decision Support Systems help at staffing secure competent employees, and provide assistance in staff training and development [5, 6]. In Help Desk Automation, Decision Support Systems help customers who have problems with the company's products and services [7, 8].

During the last years, an important direction of research that was identified is the Multi-Agent Decision Support Systems. Agents are designed to be autonomous problem-solvers, possibly communicating with other agents and users, and are therefore equipped with sufficient cognitive abilities to reason about a domain, make certain types of decisions themselves, and perform the associated actions. Agents were integrated into the Decision Support Systems for the purpose of automating more tasks for the user, enabling more indirect management, and requiring less direct manipulation of the Decision Support Systems. Specifically, agents were used to collect information and to generate decision-making alternatives that would allow the user to focus on solutions that were found to be significant [9].

Some problems may involve agents with different perspectives working on the same goal or agents working on interdependent goals. Thus, agents may face conflicting solutions to goals during their activities. Conflicts may arise even if agents try to coordinate their tasks due to their lack of an updated and complete view of the environment state and other relevant agents. Thus, the determination of conflicting knowledge is an important issue in the development of Multiple Agent Decision Support Systems for a number of reasons [10]. First, unless such conflicts are investigated, system behaviour may be affected. The combination of conflicting judgments is likely to result in system behaviour that is not sensible. Second, the existence of conflicting judgments by multiple experts suggests that the system has been miss-specified. If the system contains conflicting knowledge, one explanation is misuse or misinterpretation of information. If the system has been miss-specified in one aspect, then it may be miss-specified in others. As a result, it is critical to determine the correctness of those specifications. Third, the existence of multiple disparate judgments is likely to result in difficulties when the system is verified and validated.

As in human organizations, agents can deal with such conflicts through negotiation [11], mediation [12] and arbitration [13] strategies. Conflict resolution techniques have to be adopted to overcome such undesirable occurrences. In this paper, we discuss the idea laid behind that agents can resolve conflicts in their knowledge using intelligent techniques specifically in Multi-Agent Decision Support Systems. The rest of the paper is organized as follows. In the next section, we discuss the concepts of Decision Support Systems and Multi-Agent Systems. We then elaborate on the notion of Multi-Agent Decision Support Systems and describe knowledge conflict and intelligence techniques for conflict resolution after that. Subsequently, we illustrate detailed discussions of these intelligent techniques followed by a summary of this paper and future research directions.

## II. BACKGROUND

### A. Decision Support Systems

Decision Support Systems comprise software systems that assist humans in making complex decisions in real-life problem domains. With the advent of the powerful computing devices over the last decade, dynamic and intelligent decision support is rapidly emerging as the new research direction in the field of Decision Support Systems. Decision-making problems in real life are characterized by complex, unstructured nature of problem domains, unpredictable outcome of decisions due to the dynamic nature of problems and information, and the potential risks associated with making an incorrect/inaccurate decision. Decision Support Systems are comprised of components for sophisticated database management ability, powerful modelling functions, and simple user interface that enable interactive queries, reporting, and graphing functions [14].

### B. Multi-Agent Systems

An intelligent agent is a computer entity that is situated in some environment and can receive information from its environment and then analyse this input information to make decisions and use those decisions to take actions in its environment by using its actuators [15]. A Multi-Agent System consists of several agents, which interact with one another using a communication language. In such systems, agents can negotiate, collaborate or even compete with one another to achieve common system delegated goals [15]. Each agent has a local view of the environment; generally it has been provided by specific operational goals, and it is known that the agent is unable to solve the system tasks alone, at least with the quality, efficiency, resources, and other constraints defined by the problem.

Using of Multi-Agent Systems for solving large and complex problems are one of the most successful and efficient solutions. In some cases using of Multi-Agent Systems (e.g. distributed problem solving systems) are the best and well-known way for solving a variety of human processes. In general, one could say that a Multi-Agent System is used in domains in which (i) Data, control and expertise is distributed (e.g. in the geographic scope, the problem is distributed), (ii) Centralized control is impossible or impractical, (iii) Subsystems of a large system require interacting with each other in more flexible manner.

### C. Multi-Agent Decision Support Systems

Decision Support Systems operating in dynamic environments should therefore adapt the decision making procedure to the current parameters and constraints of the real-time environment to assist the decision maker in reaching an accurate and effective decision. Making the correct decision can be looked upon as solving a constraint satisfaction problem given the relevant historical information and a set of parameters describing the current environment. For complex applications, the solution of this problem can become quite involved. Therefore, it is difficult, and at times even impossible, for humans to make correct decisions without any computational aid. Software agents provide a suitable paradigm for automating complex tasks and solving complex problems more accurately and rapidly than humans. Software agents can enhance traditional Decision Support Systems by rapidly updating and using knowledge and domain information from a Decision Support Systems so that the agents can respond efficiently and accurately to user queries.

Agents can also be adapted to provide support for strategic decision-making and/or semi-structured problem solving. For example, a software agent can be programmed to dynamically learn system parameters and use these parameters to improve or evolve its actions so that it can reach its goal more efficiently. However, the knowledge that an intelligent agent acquires during execution cannot be stored beyond its lifetime. Discarding this knowledge would also be inappropriate as later decisions might require the experience gained by previous agents. An intelligent agent is frequently augmented with a knowledge base to store the experiences it acquires from the environment [16].

### D. Knowledge Conflict

Nycz [17] defined sources of knowledge conflicts, as follows: (i) The fight for managing specific resources. A conflict appears, when first side of the conflict is considered, that the second side of conflict should not has knowledge about a given resource, instead the second side of conflict is considered, that it such knowledge should has, (ii) Ideological conflict. It occurs when the parties to the conflict have different beliefs on the subject. These beliefs may arise, for example, with the kind of environment of system works or with adopted algorithms, (iii) Requiring the integration of various elements of the system. If there is a need to integrate some elements of the system in one unit, it's naturally a conflict occurs (i.e. different structures of knowledge, different types of knowledge representation), and (iv) Conflicts resulting from direct knowledge management system. A conflict occurs when each party considers, that it should manage the knowledge accumulated in the system, because it has the current and consistent status of this knowledge.

### E. Conflict Resolution Strategies

In the past two decades, researchers have developed various conflict resolution strategies for Multi-Agent Systems and their applications. Adler et al. [18] illustrated and experimented with eight conflict resolution strategies in network management problems. In addition, Bond and Gasser summarized eleven approaches for reconciling disparities [19]. Here we will briefly introduce the most popular conflict resolution strategies, namely, negotiation, mediation, and arbitration.

Negotiation is the most popular conflict resolution strategy for Multi-Agent Systems. Various techniques have been developed for negotiation and their existence reflects the rich diversity of humans' negotiation behaviour under different contexts [20]. Game theory-based negotiation developed by Rosenschein and Zlotkin is a typical example [21]. It is assumed that all agents are rational and intelligent, which means they make decisions consistently to pursue their own goals. It is also assumed that each agent's objective is to maximize its expected payoff which is measured by a utility scale. Utility-based negotiation [22] is an iterative conflict resolution process including the sub-processes of generating potential solutions, evaluating proposed solutions, and modifying failed proposals. The compromised solution is generated based on multi-attribute utility theory: minimizing the difference of agents' opinions and maximizing the payoff an agent expects.

Arbitration and mediation are processes in which conflicts are arbitrated or mediated by a third party. This third party does not have absolute power to modify conflicting agents' behaviours. The difference between arbitration and mediation is that, in arbitration, the decision of the third party (arbitrator) must be accepted by conflicting agents. Usually an arbitrator is equipped with the authority, more complete knowledge and more solution-search capabilities than other agents involved in the dispute [23, 21].

*F. Conflict Resolution Process*

Conflict Resolution is the process whereby two (or more) individual agents with conflicting interests reach a mutually beneficial agreement on a set of issues. The generic conflict resolution process includes conflict detection, search for solutions, and communication among agents to reach agreement with regard to the solutions that will be pursued [24]. The negotiation process can start either explicitly, when the conflict results from an agent's critique of another agent's proposal, or implicitly when separately generated proposal are inconsistent. Conflict detection in the case of Multi-Agent Systems is a complex and distributed problem, particularly in cases where agents are fully distributed and have no global view of problem solving. In these situations, all information relevant to the conflict must be explicitly communicated. Each agent's view of the current system state is potentially fragmented, it may be inconsistent with other agent's views, and it may be out-of-date. Recognition of a conflict may involve chains of inferences through multiple agents under uncertain conditions. Once a conflict or multiple conflicts are detected and possibly propagated, the issue of how to solve them arises. In the following section we discuss different intelligent techniques to solve such conflicts.

III. INTELLIGENT TECHNIQUES FOR KNOWLEDGE CONFLICT RESOLUTION

1) *Bayesian Network Based Technique:* Bayesian Network Bases Technique is based on utilizing the Bayesian framework to update an agent's belief about its opponents. It is assumed that agents have some prior knowledge about the opponent's preference. When agents need to decide, they try to find the most favourable offer and if the proposal is rejected then the agents update their knowledge.

2) *Case-Based Reasoning Based Technique:* Case-Based Reasoning Technique uses a case-based reasoning approach for negotiations in which agents make offers based on similarity of the negotiation context (including issues, opponents, and environment) to previous negotiations. Once a negotiation case is selected as the most relevant to the current negotiation, the agent might revise or adapt this case in order to meet any count-offer from the counterpart. Successful negotiation cases are kept in the case base for reuse in later negotiation case retrieval.

3) *Expert System Based Technique:* Exper System based Technique uses the intelligent expert system shell, to develop dispute avoidance ontology and software for negotiation planning systems. It is suggested that intelligent negotiation technology may add to alternate dispute resolution techniques and further diminish litigation.

4) *Fuzzy System Based Technique:* Fuzzy System Based Technique is based on a fuzzy constraint framework. In this framework, an agent, say the buyer, first defines a set of fuzzy constraints and submits one of them by priority from the highest to lowest to the opponent, say a seller, during each round. The seller either makes an offer based on the constraints or lets the buyer relax the constraints if a satisfactory offer is not available. The buyer then makes the decision to accept or reject an offer, or to relax some constraints by priority from the lowest to highest, or to declare the failure of the negotiation.

5) *Genetic Algorithms Based Technique:* Genetic Algorithms Based Technique uses genetic algorithm to resolve conflicts and considers optimization aspects. Agents use the Genetic Algorithms technique to generate a number of possible conflict free alternative situations.

6) *Ontology Based Technique:* Ontology Based Technique is based on building a shared ontology that permits agents to negotiate with most of the resolution strategies and poses fewer constraints on the agent implementation as possible. Rules, heuristics and attributes that define an agent's behaviour are placed in the ontology layer.

7) *Searching Based Technique:* Searching Based Technique uses different searching algorithms to find solutions compatible across the agents' community. Agents can generate an offer and search for agents vote to reject the offer. In another case each agent can work on some sub-problem(s) and produces proposals that represent sub-problem solutions. Proposals are integrated into shared solutions in shared memory and are critiqued by other agents.

Table 1 shows comparison of Intelligent Techniques for Knowledge Conflict Resolution.

TABLE I
COMPARISON OF INTELLIGENT TECHNIQUES FOR KNOWLEDGE CONFLICT RESOLUTION

| Technique | Strategy | Resolution Topology | Single/Multiple Conflict | Learn | Knowledge representation |
|---|---|---|---|---|---|
| Bayesian Framework [25, 26] | Negotiation | Distributed | Multiple | Yes | Bayesian Networks |
| Case-Based Reasoning [27, 28] | Negotiation | Centralized | Multiple | Yes | Cases |
| Expert System [29] | Negotiation | Centralized | Multiple | Yes | Ontology |
| Fuzzy System [30, 31] | Negotiation | Distributed | Single | Yes | Rules |
| Genetic Algorithms [32] | Negotiation | Centralized | Multiple | Yes | Flow Graph (Strings) |
| Ontology [33, 34] | Negotiation/ Mediation/ Arbitration | Centralized/Distributed | Single/Multiple | Yes | Multiple Ontologies |
| Searching [24, 35] | Mediation | Centralized/Distributed | Single | No | Data Structure |

## IV. CONCLUSION

While negotiation has been studied in other disciplines for many years, the study of negotiations of multi-agent environment is relatively new. There has been consistent evidence that using an intelligent agent to negotiate achieves better outcomes than negotiation between two human beings. In a real world Multi-Agent Systems such as in Decision Support Systems, where the agents are faced with partial, incomplete and intrinsically dynamic knowledge, conflicts are inevitable. Frequently, different agents have goals or beliefs that cannot hold simultaneously. Conflict resolution techniques have to be adopted to overcome such undesirable occurrences.

We reviewed several techniques for conflict resolution problem in Multi-Agent Decision Support Systems. We discussed Bayesian Network, Case Based Reasoning, Expert Systems, Fuzzy Systems, Genetic Algorithms, Ontological, and Searching based techniques. The choice of the specific technique for a given domain depends on the specification of the domain. For example, whether the agents are self interested, the number of agents in the environment, the type of agreement that they need to reach, and the amount and the type of information the agents have about each other.